\documentclass[a4paper,10pt,twoside]{cpc-hepnp}
\usepackage{CJK,upgreek,fancyhdr}
\usepackage{multicol}
\usepackage{graphicx}
\usepackage{booktabs}
\usepackage{amssymb,bm,mathrsfs,bbm,amscd}
\usepackage[tbtags]{amsmath}
\usepackage{lastpage}
\usepackage{multirow}
\usepackage{color}
\usepackage{float}
\usepackage{lineno}

\begin{document}
\begin{CJK*}{GB}{gbsn}

\fancyhead[c]{\small Chinese Physics C~~~Vol. 43, No. 9 (2019) 094104}
\fancyfoot[C]{\small 010201-\thepage}


\title{Energy staggering parameters in nuclear magnetic rotational bands\thanks{Supported by National Natural Science
Foundation of China (11675063, 11205068, 11847310 and 11775098) }}

\author{%
      Wu-Ji Sun (ËïÎÞ¼É)$^{1}$%
\quad Jian Li (Àî½£)$^{1;1)}$\email{jianli@jlu.edu.cn}
}

\maketitle

\address{%
$^1$ College of Physics, Jilin University, Changchun 130012, China
}

\begin{abstract}
The systematics of energy staggering for the magnetic rotational bands with $M1$ and $E2$ transition properties strictly consistent with the features of good candidates of magnetic rotational bands in the $A\sim80$, 110, 130 and 190 mass regions are presented. The regularities exhibited by these bands concerning the staggering parameter which increases with increasing spin are in agreement with the semiclassical description of shears mechanism. In addition, the abnormal behaviours in the backbend regions or close to band termination have also been discussed. Taking the magnetic dipole bands with same configuration in three $N=58$ isotones, i.e., $^{103}$Rh, $^{105}$Ag, and~$^{107}$In, as examples, the transition from chiral rotation to magnetic rotation with the proton number approaching $Z=50$ is presented. Moreover, the self-consistent tilted axis cranking and principle axis cranking relativistic mean-field theories are applied to investigate the rotational mechanism in dipole band of $^{105}$Ag.
\end{abstract}



\footnotetext[0]{\hspace*{-3mm}\raisebox{0.3ex}{$\scriptstyle\copyright$}2019
Chinese Physical Society and the Institute of High Energy Physics
of the Chinese Academy of Sciences and the Institute
of Modern Physics of the Chinese Academy of Sciences and IOP Publishing Ltd}%

\begin{multicols}{2}

\section{Introduction}

Similar to rotational bands observed in molecules, many nuclei have an energy spectrum with a pronounced rotational character and the study of nuclear rotation has been at the forefront of nuclear structure for several decades. In particular, magnetic rotation (MR)~\cite{Frauendorf1994}, which is an exotic rotational phenomenon observed in weakly deformed or near-spherical nuclei and differs from conventional collective rotation in well-deformed nuclei, has been of great interest since the observation of cascades of magnetic dipole ($M1$)
transitions in the region of neutron-deficient Pb isotopes in the 1990s~\cite{Clark1992Phys.Lett.B247,Kuhnert1992Phys.Rev.C133,Baldsiefen1992Phys.Lett.B252}.

The explanation of MR was given in terms of the shears mechanism~\cite{Frauendorf1993}. In this interpretation, the magnetic dipole vector in the magnetic rotational bands arising from proton particles (holes) and neutron holes (particles) in high-$j$ orbitals rotates around the total angular-momentum vector. Meanwhile, with increasing spin, the proton particles (holes) and neutron holes (particles) in the high-$j$ orbitals align along the total angular momentum and this alignment reduces the perpendicular component of the magnetic dipole moment. As a result, a typical property of these bands is the decreasing of the $B(M1)$ values with increasing spin.
In all, experimental indicators for magnetic rotational bands can be summarized as follows~\cite{Clark2000,Frauendorf2001,Hubel2005Prog.Part.Nucl.Phys.1,Meng2013,Meng2016}:
1) a $\Delta I = 1$ sequence of strong magnetic dipole ($M1$) transitions, corresponding to a reduced transition probability $B(M1) \sim$~a few~$ \mu_N^2$, which decrease with increasing spin;
2) weak or absent quadrupole transitions, corresponding to a deformation parameter $|\beta|\lesssim 0.15$, which combined with strong $M1$ transitions results in large $B(M1)/B(E2)$ ratios, $\gtrsim 20\;\mu_N^2/(eb)^2$;
3) a smooth variation in the $\gamma$ transition energy with angular momentum;
4) a substantial moment of inertia, corresponding to the large ratio of the $\mathcal{J}^{(2)}/B(E2) \gtrsim 100\;{\rm MeV}^{-1}(eb)^{-2}$, compared with the values in well-deformed [$\sim 10\;{\rm MeV}^{-1}(eb)^{-2}$] or superdeformed [$\sim 5\;{\rm MeV}^{-1}(eb)^{-2}$] rotational bands.

The first clear evidence of magnetic rotational bands was provided by the lifetime measurements for four $M1$-bands in $^{198,199}$Pb~\cite{Clark1997}. From then on, more and more magnetic rotational bands have been observed not only in the mass region of $A\sim190$, but also in the $A\sim60$, 80, 110 and 130 regions. To date, more than 200 magnetic dipole bands spread over 110 nuclides have been observed, which have been summarized in the nuclear chart in Fig.~\ref{fig_1} based on the review on the observed MR bands~\cite{Amita2000At.DataNucl.DataTables283} and recent observations~\cite{Li2013PhysRevC.88.014317,Zheng2015,Kumar2017,Das2018PhysRevC.98.014326,Rajbanshi2018PhysRevC.98.061304(R),Chiara2001,Trivedi2012PhysRevC.85.014327,Deo2006PhysRevC.73.034313,Yao2014,Datta2008PhysRevC.78.021306,Choudhury2015PhysRevC.91.014318, Negi2010PhysRevC.81.054322,He2011PhysRevC.83.024309,Rajbanshi2016PhysRevC.94.044318,Rajbanshi2014PhysRevC.89.014315,Li2012Nucl.Phys.A34,Rajbanshi2014PhysRevC.90.024318,Negi2012PhysRevC.85.057301,Petrache0216PhysRevC.94.064309, Bhattacharjee2009,Kaim2015PhysRevC.91.024318,Zhang2011PhysRevC.84.057302,Wang2012PhysRevC.86.064302,
PhysRevC.79.067304,Garg2015PhysRevC.92.054325,Cheng2014PhysRevC.89.054309,Procter2010PhysRevC.81.054320,Auranen2015PhysRevC.91.024324,Hartley2008PhysRevC.78.054319,
Li2016PhysRevC.93.034309,Ayangeakaa2015PhysRevC.91.044327,Pai2012PhysRevC.85.064313,Zerrouki2015,Herzan2017PhysRevC.96.014301,
Steppenbeck2012Phys.Rev.C044316,Torres2008Phys.Rev.C054318,Yuan2008HyperfineInteract.49,Agarwal2007PhysRevC.76.024321}.
The blue squares in Fig.~\ref{fig_1} represent nuclides with good candidate MR bands, in which $M1$ and $E2$ transition properties are strictly consistent with the features of MR, i.e., decreasing $B(M1)$ values with increasing spin and $B(M1)/B(E2)\gtrsim 20\;\mu_N^2/(eb)^2$, including 50 bands spread over 39 nuclides. The red squares represent other MR candidate nuclides without lifetime measurements.

\end{multicols}
\vspace{0.5cm}
\begin{center}
\includegraphics[width=14cm]{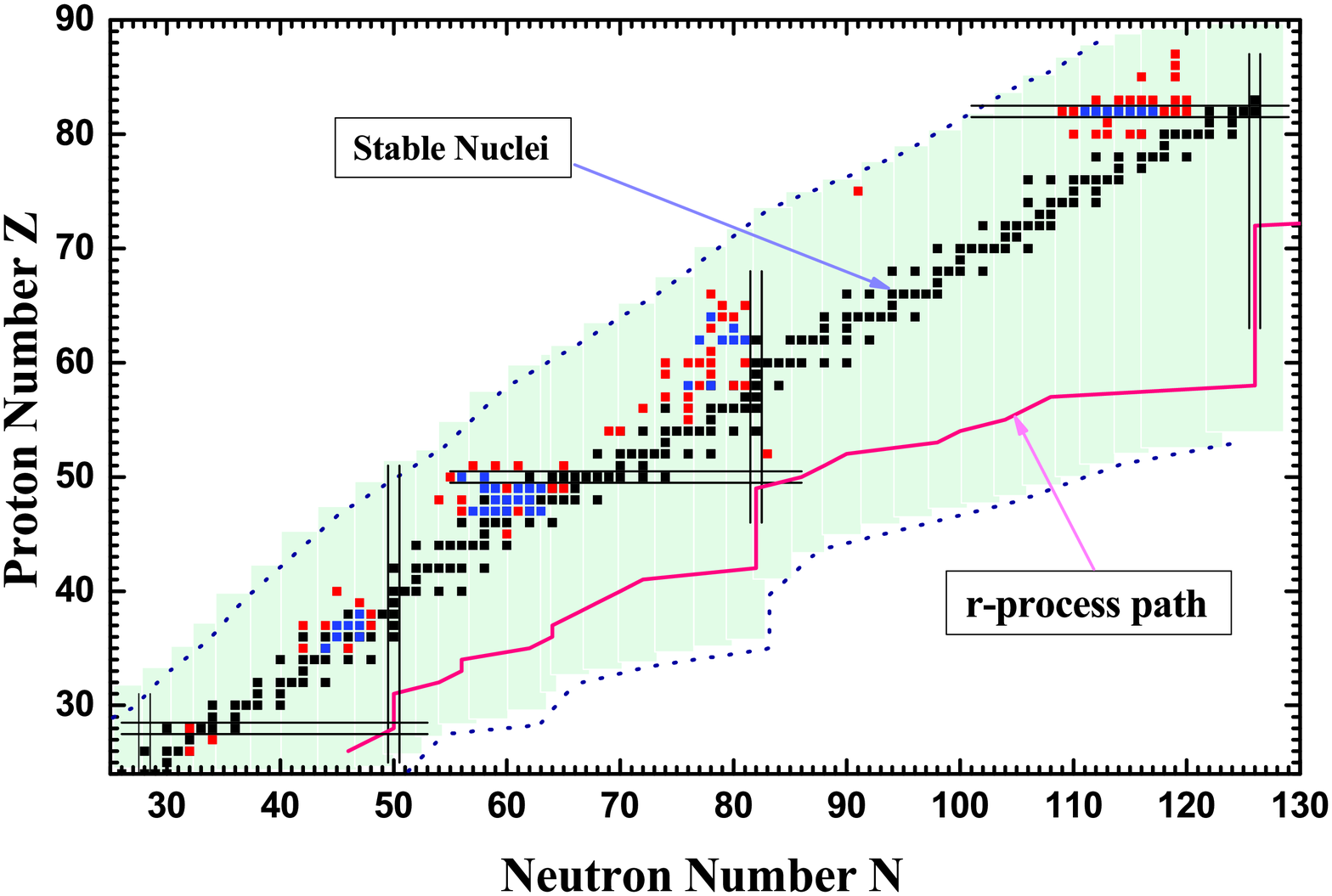}
\figcaption{(Color online) \label{fig_1} The candidate nuclides with magnetic rotation observed in the nuclear
chart. The blue squares represent nuclides with good candidate MR bands, in which $M1$ and $E2$ transition properties strictly consistent with the features of magnetic rotation. The red squares represent other candidates.
The corresponding data are taken from Refs.
\cite{Amita2000At.DataNucl.DataTables283,Li2013PhysRevC.88.014317,Zheng2015,Kumar2017,Das2018PhysRevC.98.014326,Rajbanshi2018PhysRevC.98.061304(R),Chiara2001,Trivedi2012PhysRevC.85.014327,Deo2006PhysRevC.73.034313,Yao2014,Datta2008PhysRevC.78.021306,Choudhury2015PhysRevC.91.014318, Negi2010PhysRevC.81.054322,He2011PhysRevC.83.024309,Rajbanshi2016PhysRevC.94.044318,Rajbanshi2014PhysRevC.89.014315,Li2012Nucl.Phys.A34,Rajbanshi2014PhysRevC.90.024318,Negi2012PhysRevC.85.057301,Petrache0216PhysRevC.94.064309, Bhattacharjee2009,Kaim2015PhysRevC.91.024318,Zhang2011PhysRevC.84.057302,Wang2012PhysRevC.86.064302,
PhysRevC.79.067304,Garg2015PhysRevC.92.054325,Cheng2014PhysRevC.89.054309,Procter2010PhysRevC.81.054320,Auranen2015PhysRevC.91.024324,Hartley2008PhysRevC.78.054319,
Li2016PhysRevC.93.034309,Ayangeakaa2015PhysRevC.91.044327,Pai2012PhysRevC.85.064313,Zerrouki2015,Herzan2017PhysRevC.96.014301,
Steppenbeck2012Phys.Rev.C044316,Torres2008Phys.Rev.C054318,Yuan2008HyperfineInteract.49,Agarwal2007PhysRevC.76.024321}.}
\end{center}
\begin{multicols}{2}

Signature is the quantum number specifically appearing in a deformed intrinsic system and associated with symmetry under a rotation of $180^\circ$ about the axis of nuclear rotation. A rotational band with a $\Delta I=1$ sequence could be divided into two branches classified by the signature quantum number~\cite{Bohr1975,Chen1983,Velazquez2001}. Experimentally, energy staggering between alternate spin states, which is best visualized by the experimental quantity $S(I)=[E(I)-E(I -1)]/2I$, could be observed in many rotational bands and is usually referred to as signature splitting. Theoretically, no signature splitting should be observed in ideal magnetic rotational bands due to the pure individual motion of nucleons (shears mechanism) and the tilted angle generally remains far from $90^{\circ}$ in tilted rotational mode of nuclei~\cite{Frauendorf1993,Frauendorf2001}. However, it is noted that the experimental energy staggering could be found for some MR candidates as mentioned in Ref.~\cite{Amita1999}.

In magnetic rotational bands, their energy spectra show rotational-like features with strong $M1$ transitions and it is noted in Ref.~\cite{Clark2000} that the rotational-like energy spectra  (away from band crossings) approximately follow the pattern of~$E(I)-E_0 \varpropto (I-I_0)^2$ such as in Pb isotopes~\cite{Clark1992Phys.Lett.B247,Kuhnert1992Phys.Rev.C133,Baldsiefen1992Phys.Lett.B252}, where $E(I)-E_0$ is the relative energy of a state with spin $I$ at energy $E(I)$ to the energy of the bandhead state, $E_0$. The rotational-like behaviour observed could be explained by a semiclassical analysis of the shears mechanism from a residual proton-neutron interaction~\cite{Macchiavelli1998}. Therefore in the MR bands, $S(I) \varpropto 1-(2I_0+1)/2I$ are supposed to increase with increasing spin, and signature splitting, i.e., energy staggering, is not expected. In comparison, for conventional rotation in well-deformed nuclei, its energy spectra follow the pattern $E(I)-E_0 \varpropto I(I+1)$~\cite{Bohr1975} based on the simple assumption of constant moment of inertia, and the $S(I)$ values are constant at various spins.

Thus, it is interesting to systematically study energy staggering $S(I)$ of magnetic rotational bands, especially for the ones with suitable electromagnetic transition properties. In the present work, systematic behaviours of $S(I)$ in good candidates of MR bands would be investigated. In addition, taking the dipole bands with same configuration in $^{103}$Rh, $^{105}$Ag, and $^{107}$In as examples, the characteristics of $S(I)$ in different rotational modes would be compared.

\section{Energy staggering systematics of magnetic rotational bands}

In Fig.~\ref{fig_2}, the experimental energy staggering $S(I)$ for the good candidate MR bands in the mass $A\sim80, 110, 130$, and $190$ regions are shown, which include 50 bands spread out in 39 nuclides marked with blue squares in Fig.~\ref{fig_1}. In the 110, 130 and 190 mass regions, the MR bands in even-even, odd-odd and odd-A nuclei are presented respectively. The corresponding data are taken from the previous review~\cite{Amita2000At.DataNucl.DataTables283} and recent observations~\cite{Kumar2017,Das2018PhysRevC.98.014326,Rajbanshi2018PhysRevC.98.061304(R),Chiara2001,Trivedi2012PhysRevC.85.014327,Deo2006PhysRevC.73.034313,Datta2008PhysRevC.78.021306,Banerjee2011PhysRevC.83.024316,
He2011PhysRevC.83.024309,Yao2014,Choudhury2015PhysRevC.91.014318,Negi2010PhysRevC.81.054322,Pasternak2008,Rajbanshi2016PhysRevC.94.044318,  Rajbanshi2014PhysRevC.89.014315,Rajbanshi2014PhysRevC.90.024318,Li2012Nucl.Phys.A34,Das2017}.

\end{multicols}
\vspace{0.5cm}
\begin{center}
\includegraphics[width=16cm]{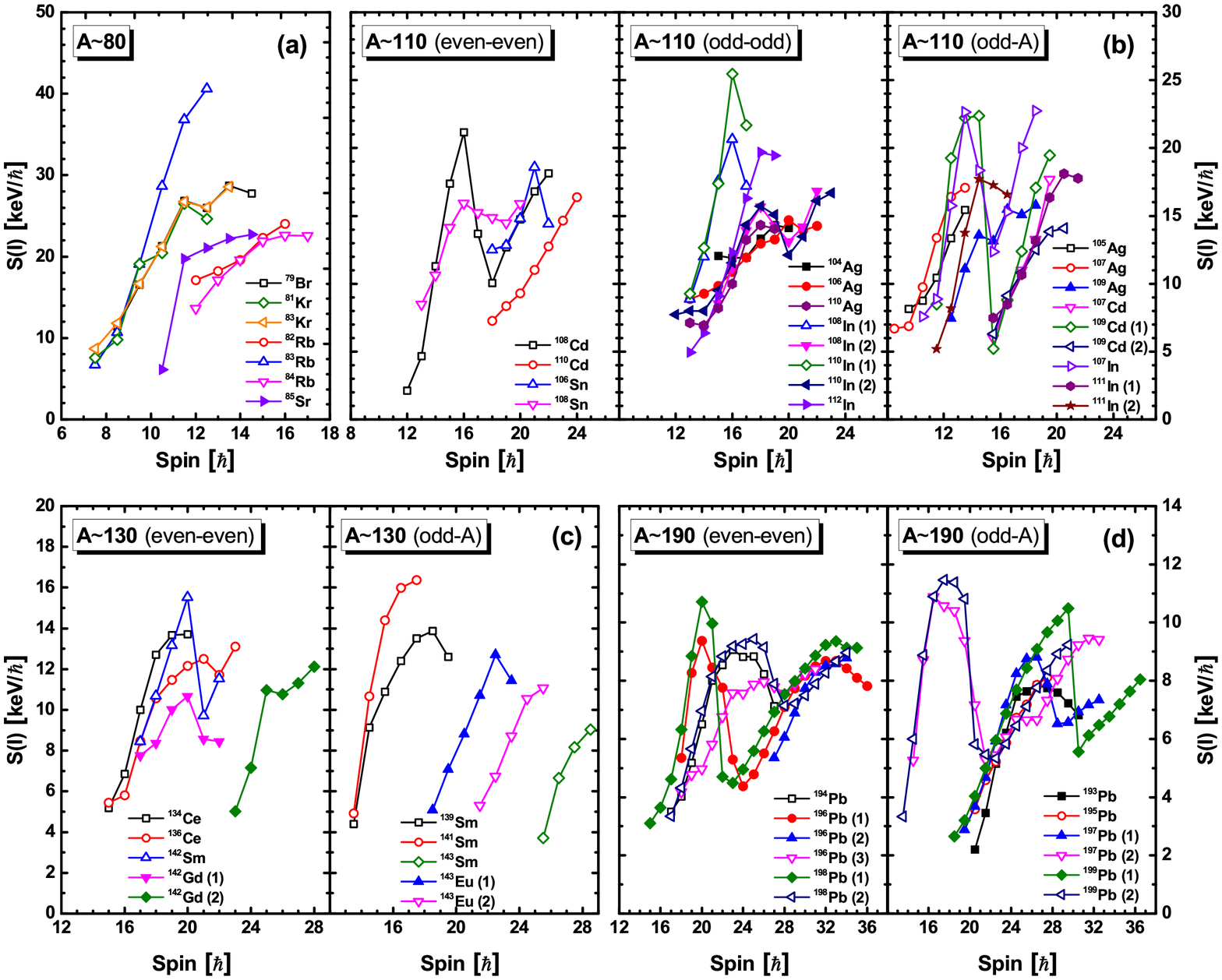}
\figcaption{
(Color online)\label{fig_2}
The experimental energy staggering $S(I)$ as a function of spin $I$ for the good candidate MR bands in the mass number $A\sim80$ (a), 110 (b), 130 (c) and $190$ (d) regions. In the 110, 130 and 190 mass regions, MR bands in even-even, odd-odd and odd-$A$ nuclei are presented separately. The filled symbols and the open symbols indicate positive parity bands and negative parity bands, respectively. The corresponding data are taken from the previous review Ref.~\cite{Amita2000At.DataNucl.DataTables283} and recent observations~\cite{Kumar2017,Das2018PhysRevC.98.014326,Rajbanshi2018PhysRevC.98.061304(R),Chiara2001,Trivedi2012PhysRevC.85.014327,Deo2006PhysRevC.73.034313,Datta2008PhysRevC.78.021306,Banerjee2011PhysRevC.83.024316,
He2011PhysRevC.83.024309,Yao2014,Choudhury2015PhysRevC.91.014318,Negi2010PhysRevC.81.054322,Pasternak2008,Rajbanshi2016PhysRevC.94.044318,
Rajbanshi2014PhysRevC.89.014315,Rajbanshi2014PhysRevC.90.024318,Li2012Nucl.Phys.A34,Das2017}. }
\end{center}
\begin{multicols}{2}

It could be seen from Fig.~\ref{fig_2} that for almost all the MR candidates, the experimental energy staggering $S(I)$ tend to increase with increasing spin except for the backbend regions (or band crossing), in agreement with the semi-classical formula for the rotational-like properties of MR, i.e. the deduced term $S(I) \varpropto 1-(2I_0+1)/2I$, and different from the semi-classical formula for conventional rotation in well-deformed nuclei. For examples, for the MR bands of $^{79}$Br, $^{81,83}$Kr, $^{82,83,84}$Rb and $^{85}$Sr in the 80 mass region, $S(I)$ clearly exhibits a increasing pattern as the spin increases. In Ref.~\cite{Macchiavelli1998Phys.Rev.C3746}, the semiclassical approach using two blades (particle and hole) interacting with an effective force is proposed and this simple scenario can account for the important features of the shears bands in neutron-deficient Pb nuclei and other mass regions. In fact, the moment of inertia for shears mechanism would decrease with the blades of the shears closing~\cite{Macchiavelli1998Phys.Rev.C3746,Frauendorf2001}. The behaviour that kinematic moment of inertia $\mathcal J^{(1)}\,(=1/[2S(I)])$ decreases with angular momentum is clearly demonstrated in Ref.~\cite{Macchiavelli1998Phys.Rev.C3746}, which further proves the increasing feature of $S(I)$ with spin. However, $S(I)$ for the MR band in $^{108}$Sn~\cite{Jenkins1999PhysRevLett.83.500} shown in Fig.~\ref{fig_2}(b) is an exception, i.e., in the higher spin region of the band, $S(I)$ begin to decrease while $B(M1)$ still shows a decreasing tendency.

It could also be seen that the $S(I)$ values in the backbend region show a decreasing trend with increasing spin for the MR bands in the 110, 130, and 190 mass regions. The backbend phenomenon is usually interpreted as the decoupling of a pair of particles from the rotating nuclear core and the subsequent rotational alignment of their angular momenta along the rotation axis~\cite{Stephens1972}. The backbend would result in increasing moment of inertia caused by the shears opening due to the gradual alignment of the particles, and therefore $S(I)$ shows the decreasing trend in the backbend regions of MR bands. For example, the $S(I)$ of negative-parity MR band of $^{199}$Pb decreases from 11.5 to 5.3 $\mathrm{keV}/\hbar$ in the backbend region~\cite{Neffgen1995Nucl.Phys.A499} as shown in Fig.~\ref{fig_2}. Here, the configurations of the negative-parity MR band in $^{199}$Pb are $\pi(h_{9/2}i_{13/2})_{K=11^-}\otimes \nu i_{13/2}^{-1}$ before backbend ($I\leq17.5\hbar$) and $\pi (h_{9/2}i_{13/2})_{K=11^-}\otimes \nu i_{13/2}^{-3}$ after backbend ($I\geqslant23.5\hbar$) respectively~\cite{Neffgen1995Nucl.Phys.A499}. After the rotational alignment of two $i_{13/2}$ neutron holes, the shears open up to $90^{\circ}$ coupling and a new shears band starts to build up.

In addition, it is easy to see in Fig.~\ref{fig_2} that the $S(I)$ values in higher mass regions are relatively smaller compared to that in lower mass regions. In the $A\sim80$ region, $S(I)$ varies from 8 to 42 $\mathrm{keV}/\hbar$ and it varies from 2 to 25 $\mathrm{keV}/\hbar$ in the $A\sim110$ region. Then in the $A\sim130$ region, $S(I)$ varies from 4 to 16 $\mathrm{keV}/\hbar$ and finally it varies from 2 to 12 $\mathrm{keV}/\hbar$ in the $A\sim190$ region. This behaviour is closely related to the mass dependence of the moment of inertia for these bands discussed in Ref.~\cite{Macchiavelli1998}. Moreover, $S(I)$ is also related to the corresponding valence particle-hole configuration. For example, for the positive-parity band in $^{108,110,112}$In~\cite{Chiara2001,Trivedi2012PhysRevC.85.014327} with the same configuration $\pi g_{9/2}^{-1}\otimes \nu h^2_{11/2}(g_{7/2}/d_{5/2})^1$, the $S(I)$ values of these bands are quite similar as shown in Fig.~\ref{fig_2}(b). While for the two bands with different configurations in the same nuclide $^{108}$In, there is a clear discrepancy in the $S(I)$ values.

However, some MR bands in Fig.~\ref{fig_2} show abnormal behaviours of $S(I)$ for the band termination (high spin) region.
For example, a sudden decline in $S(I)$ happens at the highest spin ($I=23.5\hbar$) of MR band of $^{143}$Eu in Fig.~\ref{fig_2}(c), or  $S(I)$ decreases in the high-spin region ($I>26.5\hbar$) of $^{193}$Pb in Fig.~\ref{fig_2}(d). Similar behaviours could also be seen in $^{104}$Ag, $^{110,111,112}$In, $^{106}$Sn, $^{136}$Ce, $^{139}$Sm, $^{142}$Gd and $^{194,196,198}$Pb.
The decreasing tendency of $S(I)$ in the high-spin region of the bands, in some cases a sudden decline at the highest spin, are mostly caused by the abrupt change of the configuration, which is similar to the pattern of $S(I)$ in the backbend region. While for the MR band of $^{109}$Ag in Fig.~\ref{fig_2}(b), $S(I)$ shows a staggering pattern in the high-spin region ($I\geqslant15.5\hbar$), which could also be found in $^{79}$Br, $^{81,83}$Kr, $^{105,106,107}$Ag and $^{136}$Ce as shown in Fig.~\ref{fig_2}. In Ref.~\cite{Afanasjev1999}, it is noted that the competition and interaction with collective rotation happen in the high-spin region of the bands with certain configuration, as aligned states don't have the maximum spin.

In all the presented MR bands, the increasing tendency of energy staggering $S(I)$ with increasing spin before and after backbend are observed. The systematic study of energy staggering parameter in MR bands shows a common behavior of $S(I)$, implying that $S(I)$ could be a potential indicator for MR.
Therefore, it is necessary to test and compare with other exotic rotations, such as chiral rotation~\cite{Frauendorf1997Nucl.Phys.A131}.

\section{Magnetic rotation in the $A\sim110$ region}

\subsection{The transition from chiral rotation to magnetic rotation}

Taking the $M1$ bands in $^{103}$Rh~\cite{Kuti2014PhysRevLett.113.032501}, $^{105}$Ag~\cite{Deo2006PhysRevC.73.034313,Jerrestam1995PhysRevC.52.2448,Timar2007PhysRevC.76.024307}, and $^{107}$In~\cite{Negi2010PhysRevC.81.054322,Sihotra2009} with the same configuration $\pi g_{9/2}^{-1}\otimes \nu h^1_{11/2}(g_{7/2}/d_{5/2})^1$ as examples, the extracted experimental $S(I)$ values are shown and compared in Fig.~\ref{fig_3}.

\end{multicols}
\vspace{0.5cm}
\begin{center}
\includegraphics[width=16cm]{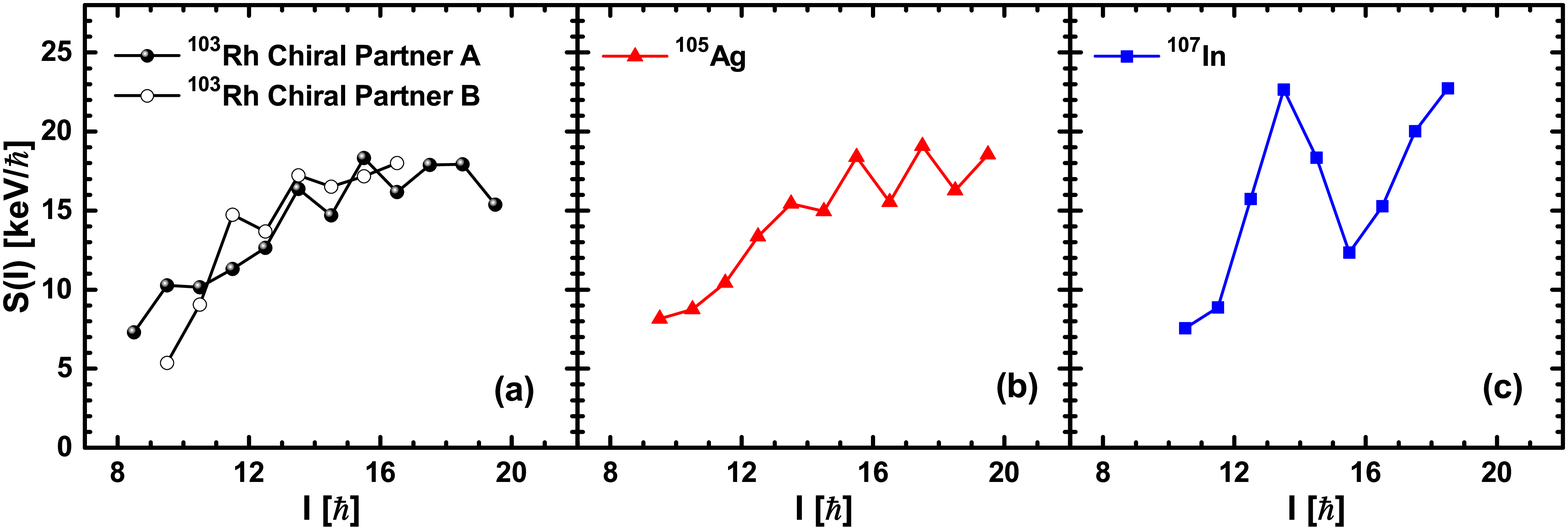}
\figcaption{
(Color online) \label{fig_3}
Experimental~$S(I)$ as a function of spin $I$ for the negative parity $M1$ bands of $^{103}$Rh (a), $^{105}$Ag (b), and~$^{107}$In~(c) with the same configuration~$\pi g_{9/2}^{-1}\otimes \nu h^1_{11/2}(g_{7/2}/d_{5/2})^1$. The corresponding data are taken from Refs.~\cite{Kuti2014PhysRevLett.113.032501,Deo2006PhysRevC.73.034313,Jerrestam1995PhysRevC.52.2448, Timar2007PhysRevC.76.024307,Negi2010PhysRevC.81.054322,Sihotra2009}.}
\end{center}
\begin{multicols}{2}

The two $M1$ bands of $^{103}$Rh have been proposed as chiral doublet bands~\cite{Kuti2014PhysRevLett.113.032501}. It should be noted that multiple chiral doublet bands with the same configuration $\pi g_{9/2}^{-1}\otimes \nu h^1_{11/2}(g_{7/2}/d_{5/2})^1$ were observed in $^{103}$Rh, and only the "yrast" chiral doublet bands are presented here. The $S(I)$ values of those two chiral partner bands in Fig.~\ref{fig_3}(a) stay almost constant with a little staggering, and $B(M1)/B(E2)$ ratios show typical values and behaviours of chiral doublet bands as suggested in Ref.~\cite{Kuti2014PhysRevLett.113.032501}.

For the $M1$ band of $^{107}$In, the $S(I)$ values increase with increasing spin before and after backbend. The behaviours of the experimental data in this band, including $B(M1)/B(E2)$ ratios and $B(M1)$ values, are consistent with ideal MR bands as suggested in Ref.~\cite{Negi2010PhysRevC.81.054322}.

For the $M1$ band of $^{105}$Ag, the $S(I)$ values of $M1$ band tend to increase with increasing spin in the lower spin region, which is consistent with the behaviour in ideal MR bands. In the higher spin region ($I>15.5\hbar$), $S(I)$ shows noticeable staggering. The staggering of $S(I)$ is usually considered as a sign of the collective rotation, which indicates a competition between shears mechanism and collective motion.

The single-particle Routhians for the proton at the top of $\pi g_{7/2}$ shell is close to the $\pi g_{9/2}$ proton shell. Considering the contribution from the collective motion is increasing, there could be a mixing of the configuration $\pi g_{9/2}^{-1}\otimes \nu h^1_{11/2}(g_{7/2}/d_{5/2})^1$ and the configuration $\pi g_{7/2}^{1}\otimes \nu h^1_{11/2}(g_{7/2}/d_{5/2})^1$ at high-spin states.

Based on the above statements, it could be concluded that the shears mechanism does not seem to dominate in the $M1$ bands of $^{103}$Rh and $^{105}$Ag, and there is an obvious transition from chiral rotation to magnetic rotation in the $A\sim110$ region when the proton number is approaching $Z=50$.
Moreover, the quadrupole deformation in previous TAC calculations are 0.26, 0.19 and 0.12 respectively for $^{103}$Rh~\cite{Kuti2014PhysRevLett.113.032501}, $^{105}$Ag~\cite{Datta2003PhysRevC.67.014325} and $^{107}$In~\cite{Negi2010PhysRevC.81.054322} with the same configuration $\pi g_{9/2}^{-1}\otimes \nu h^1_{11/2}(g_{7/2}/d_{5/2})^1$, probably indicating that the deformation of nuclei becomes smaller with increasing Z number towards $Z=50$. Similarly to Refs.~\cite{He2010PhysRevC.81.057301,Yao2014}, the increasing importance of the shears mechanism and decreasing contribution of collective rotation with increasing proton number towards $Z=50$ in the $A\sim110$ region also have been presented and discussed. Thus the systematic study of the different rotational modes in the $A\sim110$ region is an interesting question.

\subsection{The magnetic dipole band in $^{105}$Ag}

During the past few decades, relativistic mean-field (RMF) theory has been a great success in describing properties of nuclei and many nuclear phenomena~\cite{Ring1996Prog.Part.Nucl.Phys.193,Vretenar2005Phys.Rep.101,Meng2006Prog.Part.Nucl.Phys.470}. The principle axis cranking relativistic mean-field (PAC-RMF) theory has been used to describe collective rotational motion in deformed nuclei~\cite{Koepf1989Nucl.Phys.A61}. Based on the RMF theory, the tilted axis cranking relativistic mean-field (TAC-RMF) theory has been developed for describing the nuclear magnetic and antimagnetic rotational modes~\cite{Meng2013,Meng2016}. The cranking RMF model with arbitrary orientation of the rotational axis, i.e., three-dimensional cranking, has been developed and applied for the magnetic rotation in $^{84}$Rb~\cite{Madokoro2000Phys.Rev.C061301(R)}. Recently, the three-dimensional TAC-RMF theory with point-coupling interaction has been used to investigate multiple chirality in nuclear rotation~\cite{Zhao2017Phys.Lett.B1,Zhao2019PhysRevC.99.054319}. The two-dimensional cranking RMF theory based on the meson exchange~\cite{Peng2008Phys.Rev.C024313} and the point-coupling interactions~\cite{Zhao2011Phys.Lett.B181,Zhao2012Phys.Rev.C054310} has also been established and applied successfully to describe magnetic rotation in $A \sim$ 60, 80, 130 and 190 regions~\cite{Meng2013,Meng2016}, and especially the 110 region~\cite{Zhao2011Phys.Rev.Lett.122501,Zhao2012Phys.Rev.C054310,Li2012Nucl.Phys.A34,Li2012Phys.Rev.C057305,Ma2012Eur.Phys.J.A82,Zhang2014Phys.Rev.C047302,Peng2015Phys.Rev.C044329,Sun2016ChinPhysC.40.084101}.

To further examine the rotational mechanism as well as the staggering patterns of $S(I)$ for the band in $^{105}$Ag, the TAC-RMF calculations for the lower spin region of the dipole band in $^{105}$Ag with the configuration~$\pi g_{9/2}^{-1}\otimes \nu h^1_{11/2}(g_{7/2}/d_{5/2})^1$~ and the principle axis cranking RMF (PAC-RMF) calculations for the higher spin region of the band with the configuration~$\pi g_{7/2}^{1}\otimes \nu h^1_{11/2}(g_{7/2}/d_{5/2})^1$~in different signatures have been performed.
The point-coupling interaction PC-PK1~\cite{Zhao2010Phys.Rev.C054319} was used and pairing correlations were neglected. The Dirac equation for the nucleons is solved in a three-dimensional harmonic oscillator basis and a basis of 10 major oscillator shells is adopted. The calculated rotational excitation energies, total angular momenta and $B(M1)$ values in comparison with the corresponding data~\cite{Deo2006PhysRevC.73.034313,Jerrestam1995PhysRevC.52.2448,Timar2007PhysRevC.76.024307} are shown in Fig.~\ref{fig_4}. The results of TAC-RMF calculations for lower spin region and PAC-RMF calculations for higher spin region in the $M1$ band are displayed respectively. The experimental rotational frequency can be extracted as in Ref.~\cite{Frauendorf1996}: $\hbar\omega_{\rm exp}=\frac{1}{2}[E_\gamma(I+1\rightarrow I) + E_\gamma(I\rightarrow I-1)]$.

It could be seen in Fig.~\ref{fig_4}(a) that the calculated excitation energies of the band in both TAC-RMF and PAC-RMF calculations are both in good agreement with experimental data. In Fig.~\ref{fig_4}(b), for the lower spin region, the TAC-RMF results well reproduce the data, which further supports the significant contribution from shears mechanism. For the higher spin region, the PAC-RMF results are in reasonable agreement with the data indicating that the increasing contribution from the collective motion results from the intruder configuration $\pi g_{7/2}^{1}\otimes \nu h^1_{11/2}(g_{7/2}/d_{5/2})^1$. Moreover, the experimental $I-\omega$ plot are sandwiched between the corresponding plots for the configuration $\pi g_{9/2}^{-1}\otimes \nu h^1_{11/2}(g_{7/2}/d_{5/2})^1$ and $\pi g_{7/2}^{1}\otimes \nu h^1_{11/2}(g_{7/2}/d_{5/2})^1$, further supporting the mixing of these two configurations at high-spin states.

\begin{center}
\includegraphics[width=7cm]{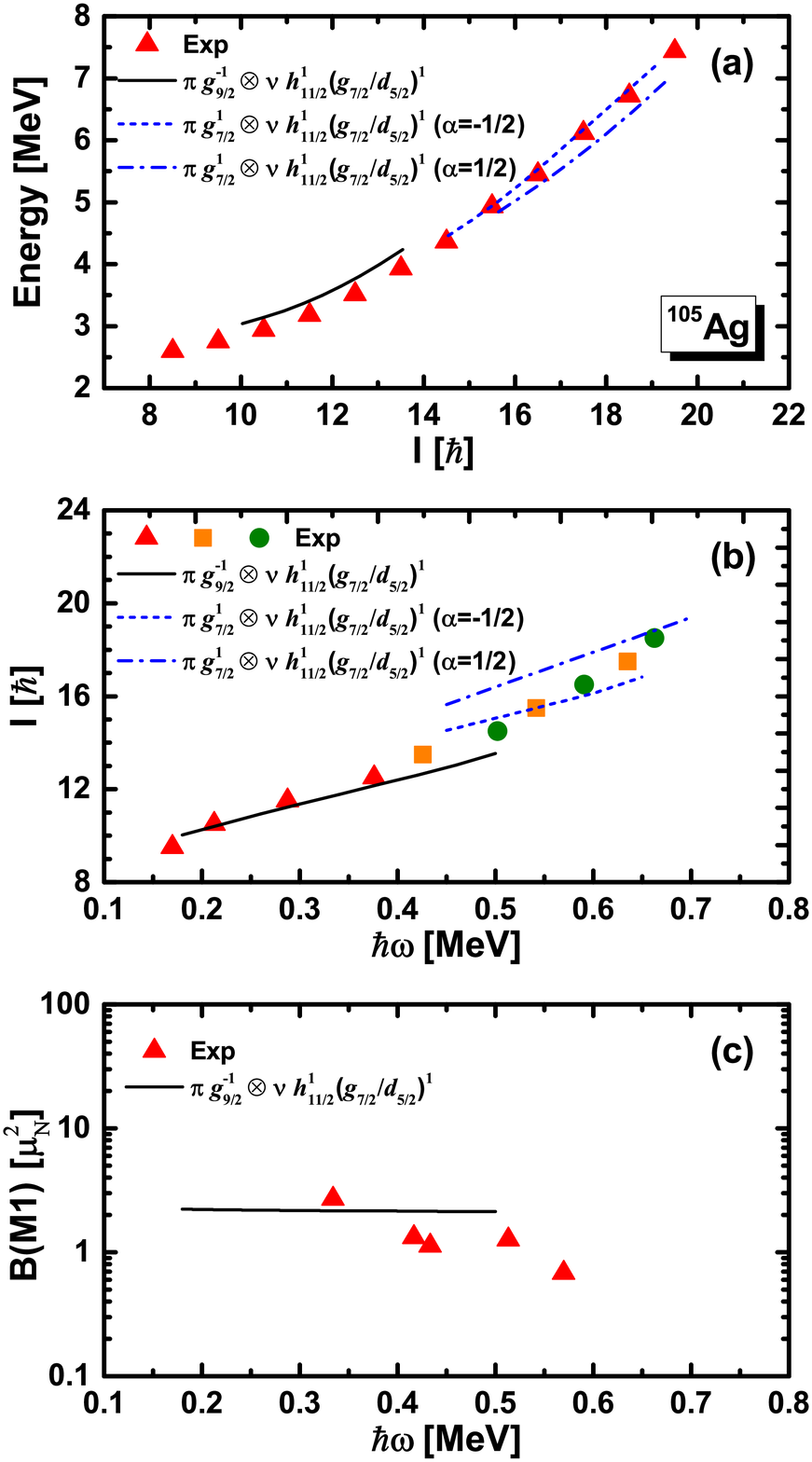}
\figcaption{
(Color online) \label{fig_4} Rotational excitation energies (a) as a function of the total angular momentum, total angular momenta (b) and $B(M1)$ values (c) as functions of the rotational frequency, for the configurations $\pi g_{9/2}^{-1}\otimes \nu h^1_{11/2}(g_{7/2}/d_{5/2})^1$ in TAC-RMF calculations and $\pi g_{7/2}^{1}\otimes \nu h^1_{11/2}(g_{7/2}/d_{5/2})^1$ in PAC-RMF calculations. The corresponding data are taken from Refs.~\cite{Deo2006PhysRevC.73.034313,Jerrestam1995PhysRevC.52.2448,Timar2007PhysRevC.76.024307}. }
\end{center}

In Fig.~\ref{fig_4}(c), the calculated $B(M1)$ values are shown to be in agreement with the data, but do not decrease much with increasing rotational frequency, which corresponds to a small decline of the shears angle.
As discussed in Ref.~\cite{Peng2008Phys.Rev.C024313}, the tilted angle $\theta$ of the orientation of the angular velocity with respect to the principal axis of the density distribution are determined self-consistently in the TAC-RMF calculations. With the rotational frequency increasing from 0.18 to 0.50 MeV$/\hbar$, the tilted angle of proton angular momentum $\theta_\pi$ changes from $5^\circ$ to $10^\circ$, the tilted angle of neutron angular momentum $\theta_\nu$ decreases from $81^\circ$ to $71^\circ$. The shears angle between these two blades decreases from $76^\circ$ to $61^\circ$, i.e., the proton and neutron angular momenta align toward each other with increasing rotational frequency, which exhibits a clear shears mechanism. The shears angle only decreases by a small amount due to the relatively high contribution from collective motion. In general, the TAC-RMF calculations support the MR interpretation for the lower spin region of the band.

\section{Conclusions}

In summary, 50 bands spread over 39 nuclides with $M1$ and $E2$ transition properties strictly consistent with the features of MR have been selected and a systematic study of energy staggering parameter $S(I)$ in these bands has been performed. The present study shows that $S(I)$ values increase with increasing spin for all the bands before and after backbends, which could be explained by the simple semiclassical description of shears mechanism. It could be treated as an indicator for MR and also needs more investigations. Moreover, the behaviours of $S(I)$ in the backbend regions or close to band termination have been discussed. In addition, the $M1$ bands in three $N=58$ isotones, i.e., $^{103}$Rh, $^{105}$Ag and $^{107}$In, with the same configuration $\pi g_{9/2}^{-1}\otimes \nu h^1_{11/2}(g_{7/2}/d_{5/2})^1$ are taken as examples to further examine the staggering behaviours of $S(I)$ in different rotational modes. It is suggested that there is a transition from chiral rotation to magnetic rotation with the proton number approaching $Z=50$, due to the competition between collective motion and shears mechanism. Furthermore, the TAC-RMF and PAC-RMF calculations have been performed, and the rotational modes in $^{105}$Ag is clearly shown.

\acknowledgments{
The authors would like to thank Dr. P. W. Zhao and Y. K. Wang for helpful discussions and collaboration during the completion of this work.
}

\end{multicols}


\vspace{-1mm}
\centerline{\rule{80mm}{0.1pt}}
\vspace{2mm}

\begin{multicols}{2}

\end{multicols}

\clearpage
\end{CJK*}
\end{document}